# Cryptographic Implications for Artificially Mediated Games

Thomas Meyer

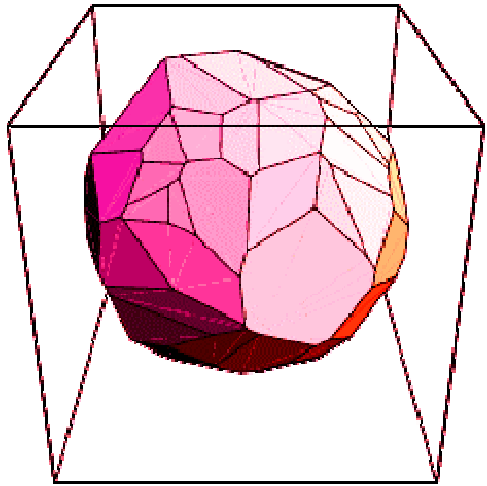

12.22.2009

Abstract

There is currently an intersection in the research of game theory and cryptography. Generally speaking, there are two aspects to this partnership. First there is the application of game theory to cryptography. Yet, the purpose of this paper is to focus on the second aspect, the converse of the first, the application of cryptography to game theory. Specifically, there exist a branch of non-cooperative games which have a correlated equilibrium as their solution. These equilibria tend to be superior to the more conventional Nash equilibria. The primary condition for a correlated equilibrium is the presence of a mediator within the game. This is simply a neutral and mutually trusted entity. It is the role of the mediator to make recommendations in terms of strategy profiles to all players, who then act (supposedly) on this advice. Each party privately provides the mediator with the necessary information, and the referee responds privately with their optimized strategy set. However, there seem to be a multitude of situations in which no mediator could exist. Thus, games modeling these sorts of cases could not use these entities as tools for analysis. Yet, if these equilibria are in the best interest of players, wouldn't it be rational to construct a machine, or protocol, to calculate them? Of course, this machine would need to satisfy some standard for secure transmission between a player and itself. The requirement that no third party (namely other players) could detect either the input or strategy profile would need to be satisfied by this scheme. Here is the synthesis of cryptography into game theory; analyzing the ability of the players to construct a protocol which can be used successfully in the place of a mediator.

I. Introduction

In the past, game theory and cryptography were developed independently of one another, deemed different both in their aims and assumptions. The first discipline models the behavior of firms and other entities in situations such as manufacturing, decisions of collusion, etc. It assumes that each party acts rationally, or in its own self-interest. Conversely, cryptography is concerned solely with secure transactions. The assumption is that the parties transacting with one another follow, to the point of irrationality, certain security protocols or cryptographic systems. The cryptanalyst, or hacker, feverishly and arbitrarily attempts to discover the content of these transactions using methods such as brute-forcing, birthday attacks, etc. Furthermore, they are believed to have access to extremely sophisticated equipment. No thought is given as to whether or not the utility of the information is greater than the resources used to illicitly obtain it. Hence, the dogmatic determination of the cryptanalyst is unfounded in certain instances.

Despite these differences, there is currently an intersection in the research of game theory and cryptography. Generally speaking, there are two aspects to this partnership. First there is the application of game theory to cryptography. Here, we use the assumptions of the former to derive new, possibly more realistic, security protocols. As the paper by Katz states:

> "Traditionally, cryptographic protocols are designed under the assumption that some parties…faithfully follow the protocol, while some parties are malicious and behave in an arbitrary fashion. The game theoretic perspective, however, is that all parties are simply rational and behave in their own best interests. This viewpoint is incomparable to the cryptographic one: although no one can be trusted to follow the protocol

(unless it is in their own best interests), the [game-theoretic] protocol need
not prevent 'irrational' behavior." [1]

That is, these protocols need not take into account what is considered as irrational behavior. For instance, if the amount of work required to break the standard is greater than the value of any data it protects, the protocol may be considered as totally secure.

This paper is primarily concerned with examining the second aspect of this union; the application of cryptography to game theory. Specifically, there exist a branch of non-cooperative games which have a correlated equilibrium (both will be described in detail in the next section) as their solution. These equilibria tend to offer more advantages compared to the more conventional Nash equilibria. Namely, they offer "better payoffs than Nash [equilibria]."[2] Additionally, these solutions "have payoffs outside the convex hull of all Nash equilibria, and therefore give more options to players."[1] The primary condition for a correlated equilibrium is the presence of a mediator within the game. This is simply a neutral entity, trusted by all parties involved in the game. It is the role of the mediator to make recommendations in terms of strategy profiles to all players, who then act (supposedly) on this advice. All transactions which the players have with this entity are, of necessity, private. Each party privately provides the mediator with the necessary information, and the mediator tells them and them alone their optimized strategy set. However, there seem to be a multitude of situations in which no mediator could exist. Correspondingly, games modeling these sorts of cases could not use these entities as tools for analysis. Yet, if these equilibria are in the best interest of players, it would be rational to simply construct a machine, or protocol, to calculate them. Of course, for this machine to be used, there would need to be some standard for secure transmission between player and machine. The requirement that no third party (namely other players) could detect either the input or strategy profile of a player would need to be satisfied by this scheme. This is the synthesis of cryptography into game theory; analyzing the ability of the players to construct a protocol which can be used successfully in the place of a mediator. Specifically, the work of Dodis, Halevi, and Rabin, in this regard will be examined. [3]

II. Basic Game Theoretic Concepts

Central to the discussion of this paper is the non-cooperative normal form game. Imagine such a game, X, is being played by $n$ players, the $i^{th}$ of which is denoted by $P_i$. Then, it is entirely described as follows: $X = \{(S_i)_{i=1}^n, (u_i)_{i=1}^n\}$, where $S_i = \{s_{i_1}, ...., s_{i_k}\}$ and $u_i = S_1 \times ... \times S_n \to \mathbb{R}$. Let $S_i$ indicate the strategy set for $P_i$, and the $s_{i_l} \in S_i$ represent the actual strategies. Hence, an outcome is simply the $n$-tuple $v = (s_{1_a}, ..., s_{n_b})$. Additionally, $u_i$ is the utility function for $P_i$; $u_i(v)$ is the utility derived from the outcome $v$ by $i$. Because this is a non-cooperative game, the following restriction is imposed: "it is assumed that each participant acts independently, without collaboration or communication with any of the others." [4]

However, non-cooperative normal form games in and of themselves are not important to this piece. Instead, their significance is derived from their ability to be extended to what are called mediated games. A mediated game, $\Psi$, is exactly as was described above. It relies on the presence of a mediator, $M$, a neutral and mutually trusted figure. The existence of *M* causes the given non-cooperative normal form game, $\chi$, to be separated into two stages. Firstly, *M* examines a known probability distribution *D* over a set of outcomes, which is dependent upon certain inputs from all the players. From this distribution, they select some vector of actions, an outcome $a = (s_1^*,...,s_n^*)$. The strategy $s_i^*$ is then recommended privately by *M* to $P_i$. As is stated by Hart and Mas-Colell "Each player is given-privately-instructions for his own play only." [5] The second stage now occurs. The game $\chi$ proceeds as before, with each player selecting some strategy from their strategy set. Of course, the convention is that each party utilizes the recommended strategy. A correlated equilibrium for $\chi$ occurs when it is rational for each entity to play their recommendation. Let $u_i(s_i', s_{-i}^* \mid k)$ be the expected utility for $P_i$ of playing $s_i' \in S_i$ as opposed to the recommendation $s_i^*$, given that all other parties play their recommendations. Then, a correlated equilibrium occurs only when $u_i(s_i', s_{-i}^* \mid k) \leq u_i(s_i^*, s_{-i}^* \mid k)$ for all players and elements of their strategy sets. This is echoed in the following theorem[*].

> **Theorem One** – Define $X = \{(S_i)_{i=1}^n, (u_i)_{i=1}^n\}$. Then, a distribution *D* leads to a correlated equilibrium if and only if for all $k = (s_{1_j},...,s_{n_t})$ created from *D* by *M*, all $P_i$, and all $s_i' \in S_i$, the condition $u_i(s_i', s_{-i}^* \mid k) \leq u_i(s_i^*, s_{-i}^* \mid k)$ holds.

III. The benefits of a Correlated Equilibrium

The concept of a correlated equilibrium was constructed by mathematician Aumann partially to address his numerous critiques of Nash equilibria. Among these critiques is a general unattractiveness of the equilibrium, strange assumptions, nearly circular reasoning, etc. He lists them succinctly as follows:

> "Now why should any player assume that the other players will play their components of such an *n*-tuple [(one of) the Nash equilibria of a game], and indeed why should they? This is particularly perplexing when, as often happens, there are multiple equilibria; but it has considerable force even when the equilibrium is unique. Indeed, there are games whose Nash equilibria appear quite unattractive even though they are unique…In a two-person game, for example, Player 1 would play his component only if he believes that 2 will play his; this in turn would be justified only by 2's belief that 1 will indeed play his component…To many this will sound like a plain old circular argument: consistent, perhaps, but hardly compelling. Nash [equilibria do] make sense if one starts by assuming that, for some specified reason, each player knows which strategies the other players are using."[6]

Regardless of the validity of these critiques, correlated equilibria do in fact offer numerous advantages over the more conventional Nash equilibria. Firstly, it has been said by many that "correlated equilibrium may be the most relevant non-cooperative solution concept." [5] This characteristic is derived from an alternative, but equivalent formulation of the equilibrium. Namely, a game Γ, an extension of X, is considered in which there are *n* players, each of whom receives a private signal (which does not necessarily originate from a mediator) at the game's beginning. These signals then serve as the basis for a player's actions, and the game continues as normal. The Nash equilibrium derived for Γ is simply the correlated equilibrium for X. Yet, this calculation is much simpler than finding the Nash equilibrium for X. This is because "with the exception of well-controlled environments, it is hard to exclude a priori that…signals are amply available to the players, and thus find their way [easily] into a [Nash] equilibrium [for Γ]." [5] That is, the Nash equilibrium for Γ almost naturally arises (not so for X), from which the correlated equilibrium to X may be derived almost effortlessly. Hence, the correlated equilibria are quire relevant to non-cooperative games in the sense that they are simple to compute.

---
[*] A rigorous proof of this theorem can be found in [1].

Apart from an ease of computation, there are many instances in which these equilibria offer greater imputations to all players than the Nash equilibrium might: "In some games there may exist a correlated equilibrium that, for every party $P_i$, gives a better payoff to $P_i$ than any Nash equilibrium."[1] Additionally, a larger range of equilibrium strategies are afforded to each player under this paradigm. This is indicated by the nature of the topological spaces which both sorts of equilibria inhabit. The set of correlated equilibria for a game, Ω, create a convex polytope. The set of Nash equilibria, Ξ, also defines a convex polytope. However, this set is of a much more complex and restricted nature, since it consists entirely of *n*-tuples which are fixed points. Additionally, let $E_i = \{\pi_i(\ell) : \ell \in \Xi\}$ denote the set of all Nash equilibrium strategies for $P_i$. Then, "the sets $[E_1, E_2,..., E_n]$ of equilibrium strategies in a solvable game are polyhedral convex subsets of the respective mixed strategy spaces."[†][4] In this, another restriction is imposed upon the possible equilibrium strategies for a given player under the scheme of Nash equilibria. Yet, none of these constraints pertain to the correlated equilibrium or its equilibrium strategies. In brief, it offers "more options to the players."[1]

IV. The Artificial Mediator

Ostensibly, the most necessary requirement for a correlated equilibrium is that all parties involved rely upon a mediator. This referee, as was briefly stated above, must have no vested interest in the outcomes of the game; the mediator must be a mutually trusted, external entity. It is easy to think of many non-cooperative games in which the existence of such a body is not possible. Yet, if it is to the benefit of all players, since it is necessary for the favorable correlated equilibrium, wouldn't it be rational to simply construct some sort of machine to perform this task? In other words, "can the trusted mediator be replaced by a protocol [or machine] that is run by the parties themselves?"[1] This machine would provide the presence of a mediator when one couldn't be found. In this, it would provide assurance of a correlated equilibrium, or at the very least, that some of the necessary steps towards one were being made.

---

[†] A proof of this can be found in [4].

Of course, such a protocol would need to satisfy certain requirements. Chiefly, when given the appropriate input from $P_i$, the outcome it computes must yield the same expected imputations as the human mediator. Additionally, all transactions between $P_i$ and the device would need to be both private and secure; no other player could discover either the input of $P_i$, or the machine's strategy recommendation for this player, this is a non-cooperative game after all.

An Iterative Turing machine, denoted $\Pi$, is the best candidate for this artificial mediator. It can be made to satisfy the first of these requirements (shown in the next section), and through further modification, the Turing machine performs its duties securely. Additionally, one can be reassured that a correlated equilibrium can be calculated in a practical amount of time. This is because it has been calculated that such a machine can compute a correlated equilibrium for a game in polynomial time.[‡]

V. A Non-Cooperative Game with Artificial Mediator

To examine the soundness of this artificial mediator, and the ability of the players to implement such a protocol, a non-cooperative game with such a referee is considered. It should be noted that because this analysis is the realm of cryptography, the game consists of certain more cryptographic conventions, which an entirely game theoretic model would never employ. Let this game be represented by $\Theta$. Each of the $n$ players is first given a security parameter, $k$. The game itself occurs in two stages, much like a game with a human mediator, $\Psi$. To begin, there is the "cheap talk" phase. Here, the parties run the protocol $\Pi$, which computes the outcome $b = (r_1^*,...,r_n^*)$. Next, $\Pi$ privately and securely gives $P_i$ the strategy recommendation $r_i^*$. Note that any party may choose to abort the game during this stage; they may refuse to provide input to the calculating machine.

The second phase of the game simply consists of each player selecting and utilizing an element of their strategy set, not necessarily their recommendation. If it is rational for all parties to use these recommended strategies, then a correlated equilibrium at $b$ ensues. As is mentioned by Katz, "It is not hard to see that the strategy vector thus specified [$b$]…[yields] expected payoffs identical to those in the original mediated game [the correlated equilibrium vector $a$ in $\Psi$]."[1] Yet, a very large assumption is made in $\Theta$, an assumption which is unacceptable in any rigorous cryptographic analysis. Namely, it was assumed that the mediator protocol was secure and completely fair. In this sense, fair refers to the receipt of a strategy recommendation by all players which provided input to the machine. That is, all transactions between a given party and the protocol are successful. But, both security and fairness could very easily not be the case for the protocol in $\Theta$. For instance, consider a cryptanalyst (possibly a player or a body under the employ of a player) who constructs a block in the communication channel between a given player(s) and the machine. Furthermore, suppose this block functions to prevent the player(s) receiving a strategy recommendation.[§] Denote such a game by $\Theta'$.

---

[‡] Interestingly enough, Nash equilibria correspond to the complexity class, NP, on this machine; they are believed to require quasi-superpolynomial time.

Despite these adverse conditions, Dodi, Halevi, and Rabin, have demonstrated that the outcome of $\Theta'$ is still a correlated equilibrium, at least in the following special case. Let $\Theta'$ consist of two players, and assume that $P_2$ receives no output from $\Pi$. Then, as they prove,[**] there is a correlated equilibrium at which $P_1$ plays $r_1^*$ and $P_2$ utilizes the minimax profile, $m_1''$ against $P_1$. The minimax profile against the i[th] player is the action $m_{-i} \in S_{-i}$ such that $\max_{r_i^*}\{u_i(r_i^*, m_{-i})\}$ is minimized. That is, the minimax profile is an element of the strategy set of all other players, where "$P_i$ [is given] its lowest possible utility, assuming $P_i$ plays a best response [calculated by the protocol] to the strategy of the other parties."[1] This solution however does not generalize well into games with more than two players. Thus it cannot be used as a basis for the claim that a correlated equilibrium exists despite a game not being fair. In point of fact, this solution is the closest to proving this generalization, despite the fact that it only functions in the aforementioned special case.

VI. Conclusion

A good deal can be concluded about the feasibility of a mediator protocol created by the players, despite the fact that no one is certain as to whether or not correlated equilibria exist in an unfair artificially mediated game. Were they to exist, then the player created protocol would be quite advantageous and therefore feasible. Regardless of its fairness, correlated equilibria would occur, and the players would enjoy the corresponding advantages: greater payoffs, etc. Alternatively, if these solutions did not exist, apart from the special case mentioned above, there would still be impetus enough to create this protocol. However, the players involved in its creation would need to be quite cautious when approaching the fairness aspect, so as to achieve correlated equilibria. For instance, the players might consider dividing up the task of securing the protocol among themselves such that no one player knows enough about the scheme to create unfairness. Additionally, the channels connecting the parties to the machine would most likely need to involve a cryptographic system(s) widely regarded as impenetrable. These measures are equally necessary for the case in which correlated equilibria exist in some instances, and not others. That is, they are needed to guarantee equilibrium among the players. In brief, because this protocol is very capable of producing advantageous results, it is always feasible. Although, it is important to note that more effort may be required on the part of the players, depending on the existence of correlated equilibria in the general case.

---

[§] Even though it is deemed unfair, the protocol is still considered secure. An insecure protocol would imply that a third party could capture and decipher the transmissions between a player(s) and the machine. This is not necessarily the case when a machine is unfair.

[**] This is established at length in [1] and [3].